\documentclass[9pt]{article}
\usepackage{spconf,amsmath,epsfig}
\usepackage{tabularx}
\usepackage{booktabs}
\usepackage{multirow}
\usepackage{subcaption}
\usepackage[dvipsnames]{xcolor}
\usepackage{url}
\usepackage{hyperref}
\usepackage{tikz}
\usepackage{mathtools}
\usetikzlibrary{arrows.meta}
\usepackage{mathtools}
\usepackage{physics}
\usepackage{tikz}
\usetikzlibrary{quantikz}
\usepackage{soul}

\usepackage{soul}%

\title{Advantages and Bottlenecks of Quantum Machine Learning\\ for Remote Sensing}
\name{
Daniela A. Zaidenberg$^{a,1}$, Alessandro Sebastianelli$^{b,c}$, \textit{Dario Spiller}$^{c,d}$, \textit{Bertrand Le Saux}$^{c}$, \textit{Silvia Liberata Ullo}$^{b}$\thanks{
$^{1}$Corresponding author. 
\textit{Email addresses}: dzaiden$@$mit.edu (DAZ), sebastianelli$@$unisannio.it (AS), dario.spiller$@$asi.it (DS), bertrand.le.saux$@$esa.int (BLS), ullo$@$unisannio.it (SLU)}%
}%

\address{
$^{a}$ Massachusetts Institute of Technology (MIT), Boston, USA \\
$^{b}$ University of Sannio, Benevento, Italy \\
$^{c}$ European Space Agency, ESRIN, $\Phi$-lab, Rome, Italy\\
$^{d}$ Italian Space Agency, Rome, Italy\\
}

\begin{document}
%
\maketitle
\begin{abstract}
This article aims to explore the potential of current approaches for quantum image classification in the context of remote sensing. After a brief outline of quantum computers and an analysis of the current bottlenecks, it shows for the first time experiments with quantum neural networks on a reference Earth observation (EO) dataset: EuroSAT. Moreover, it establishes the proof of concept of quantum computing for EO: the models trained and run on a quantum simulator are on par with classical ones. We make the open-source code available for further developments~\footnote{QNN4EO repository: \href{https://github.com/ESA-PhiLab/QNN4EO}{https://github.com/ESA-PhiLab/QNN4EO}.}.
\end{abstract}
\vspace{-8pt}
\section{INTRODUCTION \label{sec:intro}}
\vspace{-6pt}
Quantum computers leverage quantum phenomena to manipulate information and perform computation. They are expected to play a relevant part in solving computational problems, such as integer factorization (and thus encryption), thanks to their intrinsic representation of information. They have recently gained much traction as both Google and the Jiuzhang research group have reached a quantum advantage on precisely defined problems, which is to say that their quantum devices have demonstrated the ability to solve classically intractable problems~\cite{48651, zhong2020quantum}. These notable advances encourage the use of quantum devices in a variety of fields ranging from pharmacology to Artificial Intelligence (AI), and now Earth Observation (EO).

Quantum Artificial Intelligence (QAI), in particular, is an interdisciplinary field that focuses on building quantum algorithms for improving the computational tasks of AI-based models, including sub-fields like machine learning. Quantum mechanics phenomena known as superposition and entanglement allow quantum computers to perform computations in a probabilistic manner. As a consequence, QAI algorithms are expected to be much more efficient than their classical counterparts used in Computer Vision, natural language processing and robotics, even if the  entire concept of quantum-enhanced AI algorithms is still in a conceptual research domain. Building on recent theoretical proposals, initial practical studies suggest that these concepts have the possibility to be implemented in the laboratory, under strictly controlled conditions~\cite{pub.1101409450}, and open the way to the evolution of their employment and validation. 
In EO, it appears that QAI could be specifically valuable to enhance many techniques which are now commonly used. The most straightforward advantage lies in the speedup for data processing. Indeed, most quantum algorithms are significantly more cost efficient in both query and gate complexity than their classical counterparts. For instance, searching a database classically can be done in $O(N)$ while searching it using a quantum algorithm runs in $O(\sqrt{N})$~\cite{Biamonte_2017}.
In particular, various attempts to tackle land-use / land-cover classification  recently emerged. Gwaron and Levinsky proposed Quantum Neural Networks (QNN) for multiclass classification of multispectral (Sentinel-2) images~\cite{gawron2020-MS-classif-QNN-igarss}, while Cavallaro et al. used quantum versions of an ensemble of Support-Vector Machines (SVMs) to perform land-cover binary  classification of Landsat images~\cite{cavallaro2020-igarss}.

However, quantum computation is still in a fairly developmental stage. Current quantum devices such as Noisy Intermediate-Scale Quantum (NISQ) are prone to errors due to noisy measurements. As such, there is a restriction to the number of operations that can be performed before the information stored in the quantum computer become useless. To be successful, quantum computing must be able to tackle problems relating to data input and output, internal data storage and how to do this all on NISQ Computers ~\cite{Biamonte_2017}.

This paper aims to provide a brief outline of quantum computers (in Part~\ref{sec:quantum computers}), explores existing methods of quantum image classification techniques (in Part~\ref{sec:Quantum Image Classification}), then focuses on remote sensing applications (in Part~\ref{sec:application_results}). The main contribution is two-fold. it shows for the first time experiments with quantum neural networks on a reference EO dataset: EuroSAT~\cite{helber2019eurosat}. Moreover, it establishes the proof of concept of quantum computing for EO: the models trained and run on a quantum simulator are on par with classical ones. We discuss the bottlenecks of performing these algorithms on currently available open source platforms (in Part~\ref{sec:discussion_conclusion}).
\vspace{-8pt}
\section{Preliminary Information}\label{sec:quantum computers}
\vspace{-6pt}

Let's define the basic notions of quantum computing~\cite{10.5555/1206629, 10.5555/1972505}.

Qubits are the fundamental units of information held in quantum computers. A qubit exists in a superposition of 0 and 1. The state of the qubit is expressed by equation \eqref{eqn:state_qubit}.
\begin{equation}
    \centering
    \ket{\psi} = \alpha\ket{0} + \beta\ket{1}
    \label{eqn:state_qubit}
\end{equation}
In equation \eqref{eqn:state_qubit} $\ket{\psi}$ can be viewed as a vector in a Hilbert Space where:
\begin{equation}
    \centering
    \ket{0} = 
        \begin{pmatrix}
          1\\ 
          0
        \end{pmatrix}
\end{equation}
\begin{equation}
    \centering
    \ket{1} = 
    \begin{pmatrix}
      0\\ 
      1
    \end{pmatrix}
\end{equation}
The absolute squared amplitudes of quantum state, satisfying $\abs{\alpha^2}+\abs{\beta^2}=1$, describe the probability distribution of the qubit \cite{10.5555/1206629}. Consider the state $\ket{\psi} = \sqrt{\frac{1}{3}}\ket{0}  + \sqrt{\frac{2}{3}}\ket{1}$. Here, the probability of measuring the $\ket{0}$ state of your qubit is $\frac{1}{3}$ and the probability of measuring $\ket{1}$ is  $\frac{2}{3}$.

\begin{figure}[!ht]
\centering
\begin{tikzpicture}[ line cap=round, line join=round, >=Triangle]
  \clip(-2.19,-2.49) rectangle (2.66,2.58);
  \draw [shift={(0,0)}, lightgray, fill, fill opacity=0.1] (0,0) -- (56.7:0.4) arc (56.7:90.:0.4) -- cycle;
  \draw [shift={(0,0)}, lightgray, fill, fill opacity=0.1] (0,0) -- (-135.7:0.4) arc (-135.7:-33.2:0.4) -- cycle;
  \draw(0,0) circle (2cm);
  \draw [rotate around={0.:(0.,0.)},dash pattern=on 3pt off 3pt] (0,0) ellipse (2cm and 0.9cm);
  \draw (0,0)-- (0.70,1.07);
  \draw [->] (0,0) -- (0,2);
  \draw [->] (0,0) -- (-0.81,-0.79);
  \draw [->] (0,0) -- (2,0);
  \draw [dotted] (0.7,1)-- (0.7,-0.46);
  \draw [dotted] (0,0)-- (0.7,-0.46);
  \draw (-0.08,-0.3) node[anchor=north west] {$\varphi$};
  \draw (0.01,0.9) node[anchor=north west] {$\theta$};
  \draw (-1.01,-0.72) node[anchor=north west] {$\mathbf {\hat{x}}$};
  \draw (2.07,0.3) node[anchor=north west] {$\mathbf {\hat{y}}$};
  \draw (-0.5,2.6) node[anchor=north west] {$\mathbf {\hat{z}=|0\rangle}$};
  \draw (-0.4,-2) node[anchor=north west] {$-\mathbf {\hat{z}=|1\rangle}$};
  \draw (0.4,1.65) node[anchor=north west] {$|\psi\rangle$};
  \scriptsize
  \draw [fill] (0,0) circle (1.5pt);
  \draw [fill] (0.7,1.1) circle (0.5pt);
\end{tikzpicture}
\caption{The Bloch sphere represents the complex probabilistic space in which the quantum state can exist. Gate operations rotate $\ket{\psi}$ about the Bloch sphere, changing the phase and the probability amplitudes of the qubit.} 
\label{fig:bloch_sphere}
\end{figure}
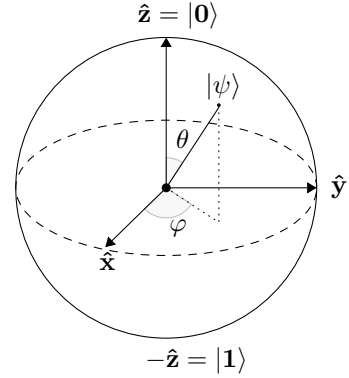

Quantum gate operations alter the phase and amplitude of qubits. Commonly used quantum operators include Pauli matrices, the Hadamard gate, the controlled NOT gate, and the $R\textsubscript{$\phi$} $ gate. Quantum gates are unitary matrices which are applied to the state vector. Single qubit gate operations can also be visualized as rotations made on the quantum state vector around the Bloch sphere,  which  represents the complex probabilistic space in which the quantum state can exist, as illustrated in Fig. \ref{fig:bloch_sphere}.

Entanglement is essential to provide the quantum advantage. By entangling two qubits, information about the state of one qubit indicates with high correlation the state of the other qubit. Moreover, the superposition property of qubits permits n qubits to describe $2^n$ possible states. These enable multiple calculations to be done simultaneously and is largely responsible for the speedup found in many quantum analogs to classical algorithms.
%
%
%
\vspace{-8pt}
\section{Quantum Image Classification}\label{sec:Quantum Image Classification}
\vspace{-6pt}
%
Like its classical counterpart, Quantum Machine Learning (QML) can be used to classify image data efficiently. In order to garner valuable information by processing data on quantum devices, hybrid quantum algorithms are necessary in order to prevent issues regarding qubit decoherence during training. Decoherence refers to the degradation of the amplitude and phase relationships of the quantum state. This can occur because of noise interference caused by several factors including minor temperature changes or vibrations~\cite{10.5555/1972505}. Without error correction, qubit decoherence causes misrepresentations in the data, rendering calculations meaningless. If qubits decohere while training, the models produced to analyze inputted data would be inaccurate. To prevent this from occurring, operations must be run on a short timescale. Thus, to train a network, data must be frequently cycled from classical to quantum states. Hybrid algorithms are a way of mediating the interactions between classical and quantum counterparts, optimizing the implementation in such a way that prioritizes run-time and model performance~\cite{phillipson2020}.
%

\begin{figure*}[!ht]
    \centering
    \includegraphics[scale=0.11]{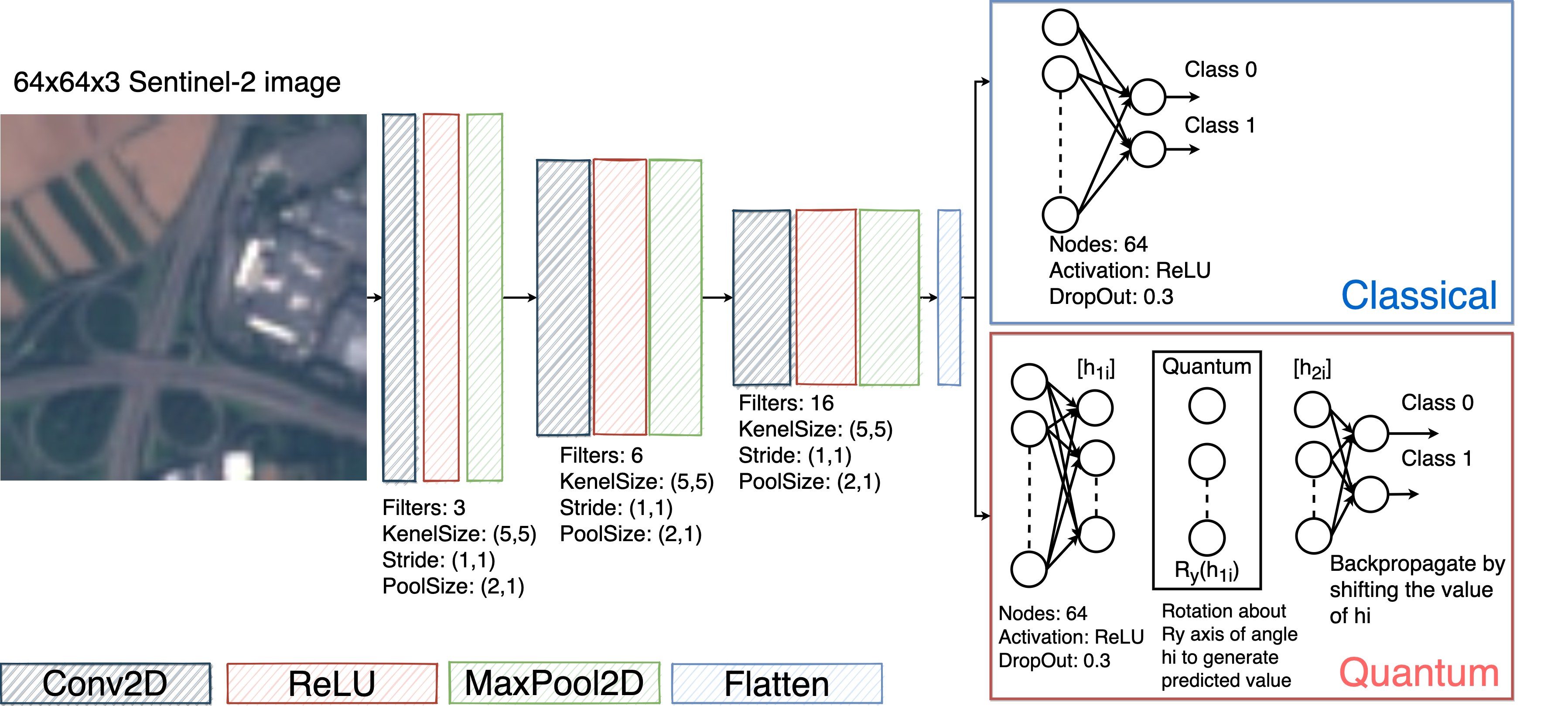}
    \caption{Proposed models: standard CNN (top); QNN4EO, a CNN with quantum embedding (bottom).}
    \label{fig:qnn_model}
\end{figure*}
%
The Quantum Neural Network (QNN) used in the experiments is a variant of the standard LeNet-5 or AlexNet presented in IBM's Qiskit Textbook~\cite{Qiskit-Textbook}. It is shown in Fig.~\ref{fig:qnn_model}, refered as QNN4EO (QNN for EO) and summarized as follows. 
The first convolutional and dense layers of the proposed QNN is identical to its classical counterpart. The convolutional branch is formed of three convolutional 2D layers followed by a ReLU and a max pooling 2D layers. 
These classical nodes are mapped to the first Fully-Connected (FC) layer of the neural network. 

In classical Convolutional Neural Networks (CNNs), it is followed by other FC layers and softmax classification.
In the quantum-classical hybrid, the classical values of the FC layer are used as the parameter to rotate some $\ket{\psi}$ along the $y$ axis of the Bloch sphere. The measurements taken after this rotation serve as the weights that connect the first classical FC layer to the last classical hidden layer. Then the final FC layer of this network is arbitrarily weighted and mapped to the output layer. Following this preliminary calculations, the process of back propagation begins. Small changes in the parametrized angle of rotation are made in the quantum layer and similar shifts are made to the classical weights as well.

It is worth noting that more complex QNNs are also being explored. In~\cite{Cong_2019}, the authors discuss a model which follows several processes of convolution and pooling. What is more, the architecture of that system works well with data classification and helps generated quantum error correcting codes for unknown error models.
\begin{figure*}
    \centering
    \includegraphics[scale=0.25]{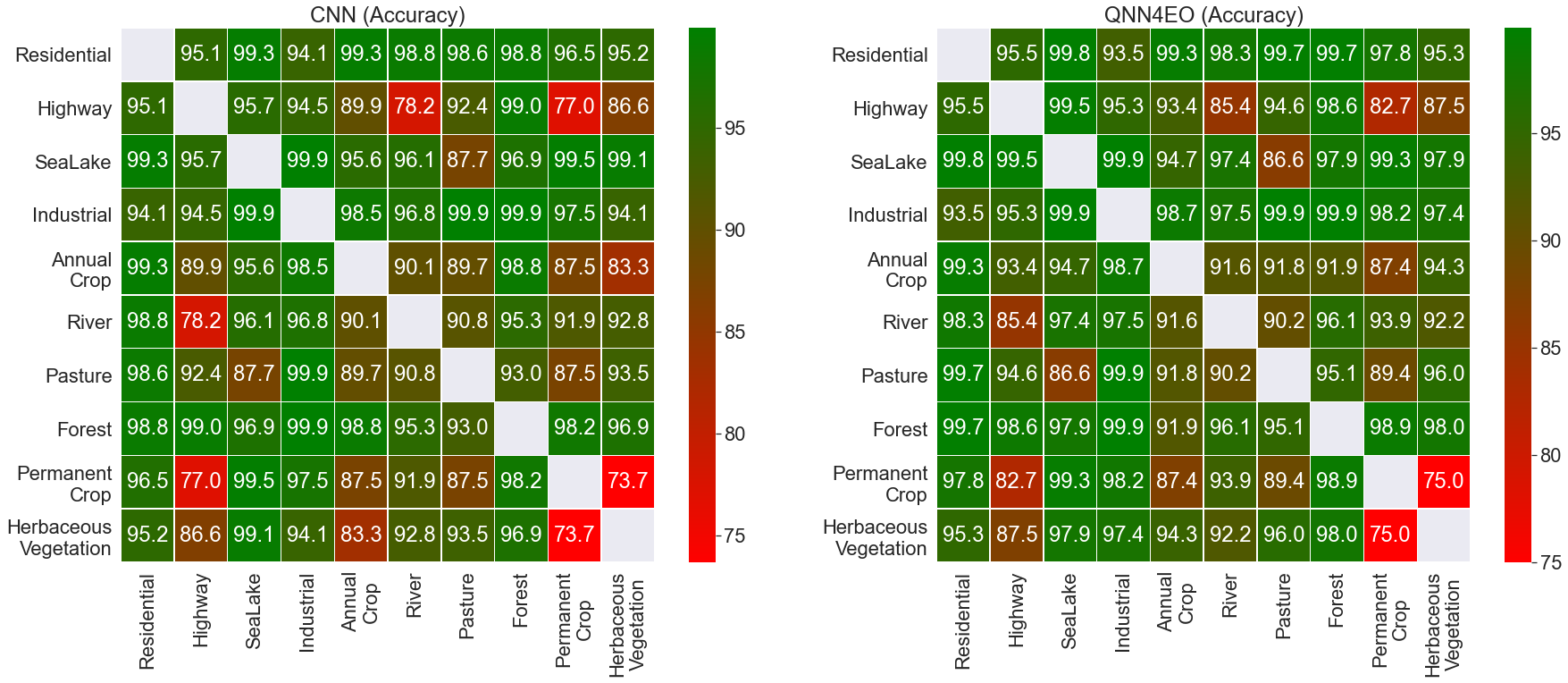}
        \vspace{-0.5em}
    \caption{Comparisons between QNN4EO and CNN}
    \label{fig:quantumVScnn}
\end{figure*}
\vspace{-8pt}
\section{Application and Results} \label{sec:application_results}
\vspace{-6pt}
The remote sensing use-case is image classification, that is identifying scenes in the EuroSat dataset~\cite{helber2019eurosat}. This dataset contains Sentinel-2 data covering $13$ spectral bands. It is divided in $10$ classes with a total of $27,000$ labeled and geo-referenced images. In order to simplify the problem, the number of classes has been reduced to two, leading to several binary classification tasks.
%
%
The dataset was then divided into training and validation with a split factor of $20\%$. Over the $13$ available bands, only the RGB ones have been selected.

The neural network presented in Part~\ref{sec:Quantum Image Classification} is implemented in a classical-quantum hybrid manner using Pytorch. The quantum component of the algorithm runs on Qiskit Aer, the high performance simulator framework provided by IBM. 
The model was trained with the backpropagation approach with one quantum node in the second hidden layer, driven by the Negative Logarithmic Likelihood loss and by the Adam optimizer, with a learning rate of $0.0001$ during $20$ epochs.

\begin{figure}[!ht]
    \centering
    \includegraphics[scale=.29]{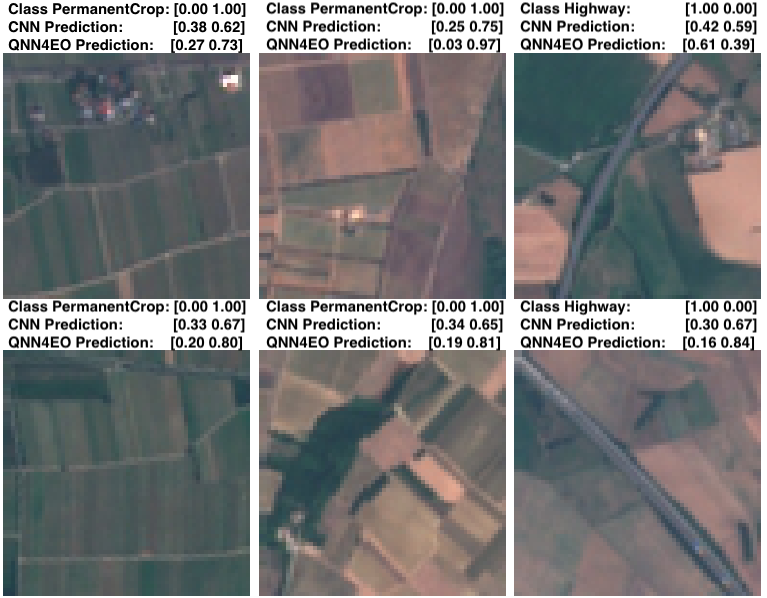}
    \vspace{-0.5em}
    \caption{Highway VS Permanent Crop}
    \label{fig:h_vs_pc}
\end{figure}

\begin{figure}[!ht]
    \centering
    \includegraphics[scale=.29]{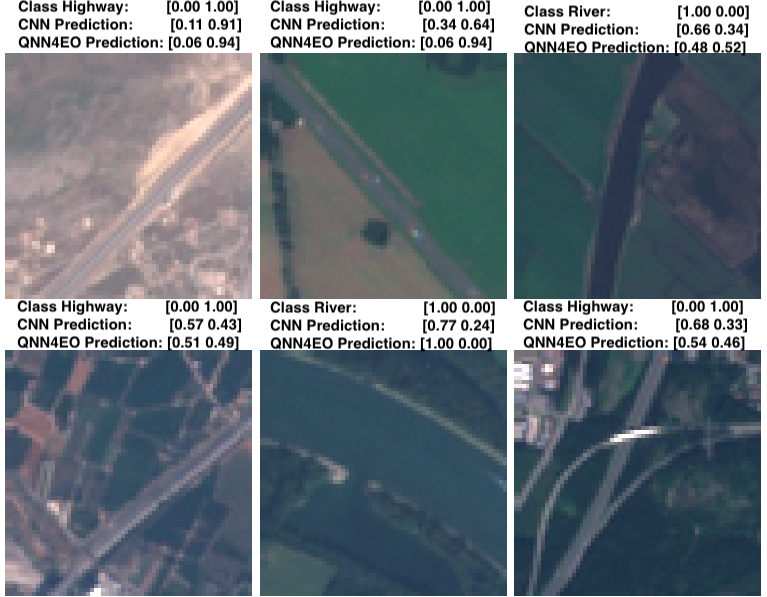}
    \vspace{-0.5em}
    \caption{Highway VS River}
    \label{fig:h_vs_r}
\end{figure}

On the validation dataset, the trained network reaches performances comparable or even greater compared to the classical CNN counterpart, establishing the  proof of concept of remote sensing image classification, as shown in Fig. \ref{fig:quantumVScnn}. 
It is worth noting that for the worst classification cases for the classical CNN (in reddish), QNN4EO slightly mitigates the learning difficulty and improve accuracy by 5 to 10\%.
In the case of Permanent Crop-Herbaceous Vegetation, the drop in performance depends on the strong similarity between the two classes. For the Highway-River and Highway-Permanent Crop cases, a possible explanation is the presence of portions of one class in images of the other class (e.g. Highway running through permanent crop areas), as shown in Fig. \ref{fig:h_vs_pc}, or similar shapes between the two classes (e.g. highway and rivers), as shown in Fig. \ref{fig:h_vs_r}.
The overall average accuracy reached by the CNN is $93.63 \%$, while the the QNN4EO reached an accuracy of $94.73 \%$. These experiments can be replicated and extended using the code available at \href{https://github.com/ESA-PhiLab/QNN4EO}{https://github.com/ESA-PhiLab/QNN4EO}.
\vspace{-8pt}
\section{Discussion and Conclusion} \label{sec:discussion_conclusion}
\vspace{-6pt}
Initial results demonstrate feasibility of QML applied to EO. Next steps include expanding the size of the quantum hidden layer and introducing a multi-class classification approach . Three challenges, very peculiar to quantum computing, are currently under investigation. First, given the size of current quantum circuits, only small-size data can be fit in and processed. This constraint is particularly limiting for image processing. So dimension reduction techniques (such as the CNN embedding of QNN4EO) are applied to encode the images before transfer them on quantum chips. Second, Qubit decoherence is still a significant hurdle in implementing most algorithms although quantum error correction is progressing. Estimating how long and how complex the data processing can be with current means is a matter of importance for porting real-life applications to quantum. Finally, reading out the outcome of the quantum process, which is essentially a probabilistic entity, requires smart sampling and statistical analysis. To harness the power of quantum computing for EO, it is essential to estimate how crucial these bottlenecks are. Even if currently not at operational status, quantum computers might become the only means to handle the EO data ever-increasing stream. 
\vspace{-12pt}
\section{Acknowledgments}
\vspace{-6pt}
Daniela A. Zaidenberg participated under a joint program of MIT and University of Sannio through the MIT Science and Technology Initiative (MISTI). This work is part of ESA $\Phi$-Lab's Quantum Computing for Earth Observation (QC4EO) initiative.
%
\vspace{-8pt}
\small
\bibliographystyle{IEEEbib.bst}
\bibliography{main}
\end{document}